\documentclass[10pt,twocolumn,aps,prx,floatfix,superscriptaddress,nofootinbib]{revtex4-2}
\usepackage{amsmath}
\usepackage{amsfonts}
\usepackage{amssymb}
\usepackage{graphicx}
\usepackage{color}
\usepackage{bm}
\usepackage{dsfont}
\usepackage{soul}
\usepackage{mathtools}
\usepackage[normalem]{ulem}

\DeclarePairedDelimiterX\braket[2]{\langle}{\rangle}{#1 \delimsize\vert #2}
\usepackage[final]{hyperref} 
\hypersetup{
	colorlinks=true,       
	linkcolor=blue,        
	citecolor=blue,        
	filecolor=magenta,     
	urlcolor=blue         
}

\newcommand{\new}[1]{\textcolor{black}{#1}}

\newcommand{\beq}{\begin{equation}}
\newcommand{\eneq}{\end{equation}}

\begin{document}

\title{Parity readout in Majorana box qubits from the dispersive to the resonant regime}

\author{Sara M. Benjadi}
\affiliation{Institut f\"ur Theoretische Physik, Heinrich-Heine-Universit\"at, 40225 D\"usseldorf, Germany}

\author{Reinhold Egger}
\affiliation{Institut f\"ur Theoretische Physik, Heinrich-Heine-Universit\"at, 40225 D\"usseldorf, Germany}

\begin{abstract} 
We study theoretical models for charge reflectometry and capacitive readout of the Majorana parity degree of freedom 
in Majorana box qubits, taking into account decoherence channels within the framework of the Lindblad master equation.  
Noting that a parity-dependent dynamical susceptibility $\chi_z(\omega)$ governs both readout schemes,
we provide a general expression for $\chi_z(\omega)$ which covers the full crossover from the resonant regime to the off-resonant
dispersive regime.  In addition, we re-examine previous results which were obtained under a semiclassical factorization assumption.
Using three different error measures, we show that this approximation is quantitatively justified in the dispersive regime.
In the resonant regime, however, we find deviations from exact reference data, obtained by numerical solution for the steady
state of the full Lindblad equation.  These deviations are typically of the order of a few percent in the considered error measures.
 \end{abstract}
\maketitle

\section{Introduction}\label{sec1}

Majorana bound states (MBSs) are predicted to exist in various platforms.  Noting that a spatially separated pair of MBSs is equivalent to a single (nonlocal) fermion state,
MBSs are of fundamental physical interest and could potentially be used as basic constituents 
in topological quantum information processing schemes. However, after initial enthusiasm \cite{Mourik2012}, the clear-cut experimental identification of MBSs and, in particular, the operation of a Majorana qubit turned out to be a rather challenging endeavor. 
Many experiments were based on proximitized semiconductor nanowires
realizing one-dimensional (1D) topological superconductors, where 
MBSs should be located at the ends of the wire.
However, a number of alternative platforms, e.g., proximitized topological insulator devices \cite{Fu2008,Hasan2010}, have also been studied.  We refer to Refs.~\cite{Lutchyn2018,Prada2020,Beenakker2020,Flensberg2021} for 
comprehensive reviews up to the year 2021.  
More recently, the so-called topological gap protocol was employed for identifying 
parameter regions where devices are expected to harbor MBSs \cite{Aghaee2023,Aghaee2025,aghaee2025a}. In addition, many aspects of Majorana physics 
can be emulated using a poor man's version of short Kitaev chains engineered by the fine tuning of coupled quantum dots
\cite{Leijnse2012,Dvir2023,tenHaaf2024,Nitsch2025,vanloo2025singleshotparityreadoutminimal}.

An intensely pursued topologically protected qubit candidate is defined by 
the Majorana box qubit (MBQ) \cite{Plugge_2017}, \emph{aka} tetron \cite{Karzig2017}.  
For the MBQ, one considers a device harboring four MBSs under strong Coulomb blockade
conditions such that the computational space associated with MBSs becomes equivalent to a qubit space \cite{Beri2012,Altland2013,Beri2013}. 
The Coulomb charging energy then provides a natural protection against quasiparticle poisoning mechanisms and allows for unique functionalities of the resulting Majorana qubit. As a consequence, several theoretical proposals for scalable architectures relying on MBQs
as basic elements have been put forward \cite{Landau2016,Plugge2016,Karzig2017,Litinski2017,Litinski2018,Aasen2025},
see also Refs.~\cite{Aasen2016,Hyart2013,Vijay2015} for related schemes.  
A key element in all these proposals is the practical availability of a reliable and fast readout scheme for measuring the parity of a MBS pair.  Using the Majorana operators $\gamma_j=\gamma_j^\dagger$ with $\gamma_j^2=1$, where different Majorana operators anticommute, the parity operator for the Majorana pair $\gamma_2$ and $\gamma_3$ in Fig.~\ref{fig1} is  $\hat z=i\gamma_2\gamma_3$ with eigenvalues $z=\pm 1$.

There are various options for performing a parity readout. 
Interferometric conductance and/or charge reflectometry parity readout schemes for MBQs have been suggested in Ref.~\cite{Plugge_2017}, see also Refs.~\cite{Ohm2015,Yavilberg2015,Vijay2016,Smith2020}.  Parity readout can also be achieved by charge sensing of a quantum dot coupled to the MBS pair \cite{Gharavi2016,Munk2019,Munk2020,Steiner2020,Szechenyi2020,Zhang2021,Schulenborg2021,Schulenborg2023}. Recent parity readout experiments  \cite{Aghaee2025,vanloo2025singleshotparityreadoutminimal} were performed by measuring the quantum capacitance $C_Q$ (which depends on $z$ as detailed below) in the dispersive regime, see also Refs.~\cite{Karzig2017,Liu2023,Sau2025}.
 Motivated by the recent experimental
 progress \cite{Aghaee2025,vanloo2025singleshotparityreadoutminimal}, we here provide 
 a theoretical analysis of parity readout in a MBQ, addressing also the crossover from the dispersive to the resonant limit.  Our results are basically agnostic to the choice of 
 Majorana platform, and may also be of interest for analyzing  
 topologically trivial cases, where MBSs are effectively replaced by Andreev bound
 states \cite{Prada2020}.  In the context of capacitive parity readout in the 
 dispersive limit, such aspects  have recently been studied for noninteracting 
 Majorana wires \cite{Sau2025}.  For concreteness, however, we here focus on the ideal MBQ case.  It is worth noting that the resonant (instead of the dispersive) regime can offer distinct advantages for MBQ parity readout \cite{Plugge_2017}.

The remainder of this paper is structured as follows.
In Sec.~\ref{sec2}, we discuss two different MBQ parity readout schemes in some detail, namely charge reflectometry and measurements of the 
quantum capacitance. In both cases, a parity-dependent dynamical susceptibility $\chi_z(\omega)$ encodes the parity sensitivity. 
To be specific, we separately study the resonant regime and the dispersive limit, where previous work \cite{Plugge_2017} has employed a
semiclassical approximation to determine $\chi_z(\omega)$ from the Lindblad master equation.  The latter equation governs the dynamics in the presence of decoherence channels. 
In Sec.~\ref{sec3}, we present a general expression for the susceptibility $\chi_z(\omega)$ which covers the entire crossover from resonant to off-resonant conditions. In addition, we examine the accuracy of the semiclassical approximation in both the strong-coupling (resonant) limit and in the weak-coupling (off-resonant dispersive) regime by comparing to exact reference results for experimentally realistic parameter choices.  Here, reference results were either obtained by numerical solution for the steady state of the full Lindblad equation (in the strong-coupling regime) or follow from exact expressions (in the weak-coupling limit).  Finally, Sec.~\ref{sec4} contains concluding remarks.  
Throughout the paper, we put $\hbar=k_B=1$.

\section{MBQ parity readout schemes}\label{sec2}

In this section, we discuss different MBQ parity readout schemes.  After summarizing the setup and the theoretical model in Sec.~\ref{sec2a},
we describe two-port charge reflectometry readout in Sec.~\ref{sec2b}.  The corresponding results in the resonant regime have already appeared in Ref.~\cite{Plugge_2017} (using the semiclassical approximation), but they are briefly reviewed here in order to keep the paper self-contained.  In the dispersive limit, we obtain exact results beyond those in Ref.~\cite{Plugge_2017}.   Next, in Sec.~\ref{sec2c}, we discuss MBQ parity readout via the quantum capacitance and show that the resonator susceptibility $\chi_z(\omega)$ appearing in the modeling of charge reflectometry
readout also determines the capacitive readout scheme.  Finally, in Sec.~\ref{sec2d}, we show that results for single-port charge reflectometry (used in Ref.~\cite{vanloo2025singleshotparityreadoutminimal}) can be directly inferred from the corresponding two-port results. 

\subsection{General remarks}\label{sec2a}

\begin{figure}
\includegraphics[width=0.5\textwidth]{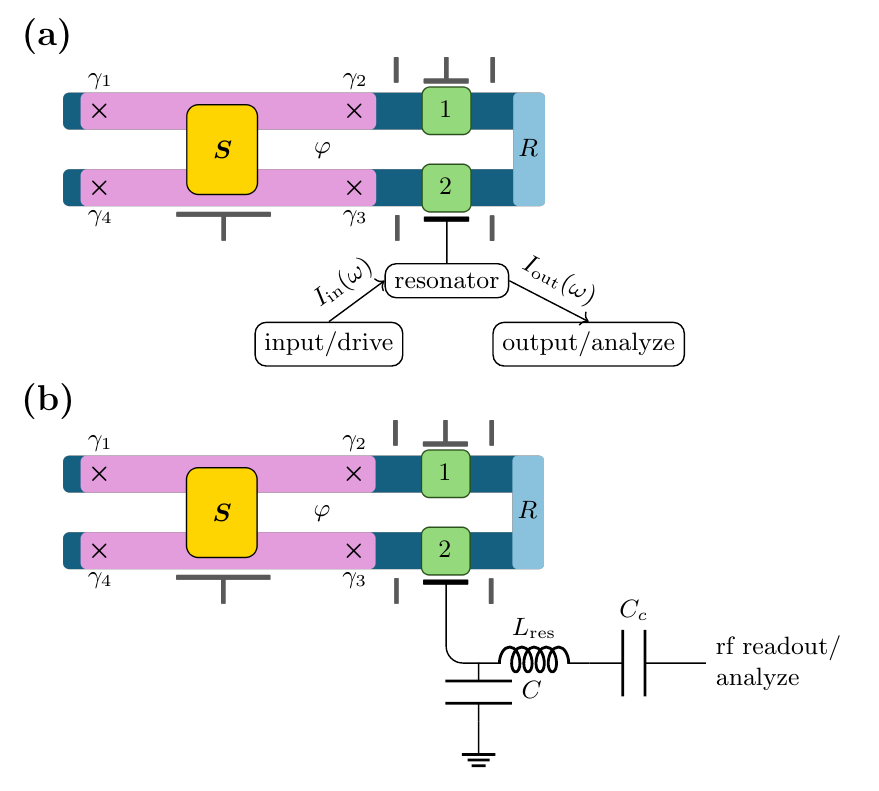}
\caption{Schematic readout schemes for the Majorana qubit parity $\hat z=i\gamma_2\gamma_3$ with eigenvalues $z=\pm 1$, see Eq.~\eqref{MBQpauli}. The Majorana states $\gamma_{1,2,3,4}$ are 
boundary states of a pair of topological superconductor wires connected by a superconducting bridge ($S$) \cite{Plugge_2017,Karzig2017}. 
The floating Majorana box is operated under Coulomb valley conditions. A pair of small quantum dots (1 and 2, shown as green squares), with the double dot in the single-electron occupancy regime, is tunnel-coupled to the wire ends harboring $\gamma_2$ and $\gamma_3$, respectively.  In addition to elastic cotunneling through the MBQ, coherent tunneling between dots 1 and 2 can proceed through a tunnel-coupled reference arm ($R$).  The phase difference $\varphi$   between both interfering tunneling paths can be tuned by gate voltages or by a weak perpendicular magnetic field. The fluctuating charge on dot 2 is monitored in the frequency domain by capacitively coupling dot 2 to a microwave cavity (photon resonator) which in turn is coupled to the readout circuit.
(a) Charge reflectometry setup, where the outgoing intensity $I_{\rm out}(\omega)$ for given intensity $I_{\rm in}(\omega)$ in the input line 
is measured in a microwave circuit. 
(b) Quantum capacitance measurement, where the cavity is used for gate sensing and capacitance readout \cite{Aghaee2025}.     }  \label{fig1}
\end{figure}

A schematic setup for MBQ parity readout is shown in Fig.~\ref{fig1}, where for concreteness, we assume a MBQ implementation in terms of two topological superconductor nanowires connected by a conventional superconductor bridge \cite{Plugge_2017,Karzig2017}.  The entire superconducting island, which constitutes a topological variant of a Cooper pair box, is kept floating and is thus characterized by a finite charging energy $E_C$. \new{For MBQs with lateral size of a few hundred nanometers, we estimate $E_C\sim 1$~meV.} Operating the superconducting island near a Coulomb valley, the charging energy enforces a fermion parity constraint for energies $|E|\ll E_C$. 
Assuming broken time-reversal symmetry, the important low-energy MBQ degrees of freedom are 
the Majorana fermion operators $\gamma_1,\ldots,\gamma_4$ in Fig.~\ref{fig1}, with the respective MBSs localized at the wire ends.   The fermion parity constraint can then be expressed in terms of the total Majorana parity as, say, $\gamma_1\gamma_2\gamma_3\gamma_4=+1$.
The constrained Hilbert space of the island defines a nonlocal qubit  with Pauli operators \cite{Plugge_2017}
\begin{equation}\label{MBQpauli}
    \hat x=i\gamma_1\gamma_2, \quad \hat y=i\gamma_3\gamma_1, \quad \hat z=i\gamma_2\gamma_3.
\end{equation}
The spatial separation of the MBS pairs defining each Pauli operator provides a topological qubit protection mechanism \cite{Beri2012,Altland2013,Beri2013,Landau2016,Plugge2016}.
 We note that, for instance, disorder effects may cause topologically trivial low-energy Andreev bound state quasiparticles \cite{Prada2020}.  However, the approaches described below also allow for the readout of Andreev state parities, see also Refs.~\cite{Aghaee2025,Sau2025}.
In what follows, we focus on the idealized case where only the $\gamma_j$ are kept as relevant degrees of freedom.

We study projective MBQ parity readout schemes for $z=\pm 1$, performed either by using charge reflectometry, see Fig.~\ref{fig1}(a), or through a measurement of the quantum capacitance, see Fig.~\ref{fig1}(b). However, by a modification of the setups in Fig.~\ref{fig1}, also the parity operators $\hat x$ and 
$\hat y$ in Eq.~\eqref{MBQpauli} can be measured \cite{Plugge_2017,Karzig2017}.
In Fig.~\ref{fig1}, two quantum dots (1 and 2) are connected through tunnel barriers 
to the MBQ via $\gamma_2$ and $\gamma_3$, respectively. The double dot is operated in the coherent single-electron occupancy regime, such that only a single electron occupies the double dot on average. \new{The corresponding dots are significantly smaller than the MBQ structure and therefore come with an even higher charging energy than the MBQ, giving direct access to the single-electron occupancy regime for the double dot \cite{Plugge_2017}. } In addition to inter-dot cotunneling processes via the MBQ with effective tunneling amplitude $t_1 \hat z$ \cite{Plugge_2017,Karzig2017}, 
we include a direct reference arm with inter-dot tunneling  amplitude $t_0$.  
The tunable phase $\varphi$ in Fig.~\ref{fig1} refers to the relative phase difference between the complex amplitudes $t_0$ and $t_1$.
The effective inter-dot tunneling amplitude due to the interference of both pathways is then given by 
\begin{equation}\label{tzdef}
    t_z=t_0+t_1 \hat z.
\end{equation}
Note that the transmission probability $|t_z|^2$ depends on the MBQ parity $z=\pm$.

In the single-electron occupancy regime of the double dot, a basis for the Hilbert space of the double dot is given by localized states, $\{|1\rangle_d,|2\rangle_d\}$, where we assume that only a single quantum level in each dot is energetically accessible. In this basis, the double dot in the presence of the tunneling amplitude $t_z$ in Eq.~\eqref{tzdef} is described by the Hamiltonian
\begin{equation}\label{hamiltonian}
    H_d = \left(\begin{array}{cc} \varepsilon & t^{}_z \\ t^\ast_z & -\varepsilon + \lambda \hat Q_s \end{array}\right),
\end{equation}
where the bare level energies $\pm \varepsilon$ are due to the (gate-tunable)
detuning $\varepsilon$ and the operator $\hat Q_s$ refers to a photon resonator (cavity) capacitively coupled to dot 2 with (weak) coupling strength $\lambda$.  For $\lambda=0$, the electron undergoes Rabi oscillations between both dots with the parity-dependent frequency
\begin{equation}\label{omegaz}
    \omega_{z=\pm} = \sqrt{\varepsilon^2+|t_0+z t_1|^2}.
\end{equation}
Transforming from $\{ |1\rangle_d,|2\rangle_d\}$ to the Rabi eigenbasis for the Hilbert space of the double dot,
we define Pauli operators $\tau_{x,y,z}$  with $\tau_\pm = (\tau_x\pm i \tau_y)/2$ with respect to the Rabi basis.

\subsection{Charge reflectometry}\label{sec2b}

Let us first address charge reflectometry readout \cite{Petersson2010,Frey2012,Colless2013,Liu2014,Plugge_2017}, see Fig.~\ref{fig1}(a), where a microwave resonator with photon creation operator $a^\dagger$ and bare frequency $\omega_0$ is capacitively coupled to the charge on dot 2, with $\hat Q_s=a+a^\dagger$ in Eq.~\eqref{hamiltonian}. For small coupling $\lambda$ in Eq.~\eqref{hamiltonian}, the Hamiltonian describing the resonator coupled to the double dot then takes the form \cite{Plugge_2017}  
\begin{equation}\label{hamfull}
    H= \omega_z \tau_z + \omega_0 a^\dagger a + \frac{\lambda}{2}
    (a+a^\dagger)\left(1-\frac{\varepsilon}{\omega_z}\tau_z -\frac{|t_z|}{\omega_z} \tau_x
    \right).
\end{equation}
We now separately discuss the limiting cases of strong coupling (near-resonant conditions) and weak coupling (dispersive off-resonant conditions).  \new{We note that in both limits, the coupling $\lambda$ remains a small parameter.}

\subsubsection{Near-resonant conditions}

First, if the resonator and the double dot are nearly resonant,  $|\omega_0-2\omega_z|\ll {\rm min}(\omega_0,\omega_z)$ for both values of $z=\pm$, the standard rotating-wave approximation (RWA) applies \cite{Blais2004}. As a result,  Eq.~\eqref{hamfull} yields
the effective Hamiltonian 
\begin{equation}
    H_{\rm RWA} = \omega_z\tau_z+\omega_0 a^\dagger a+ g_z (a\tau_++a^\dagger \tau_-),
\end{equation}
where the parity-dependent coupling
\begin{equation}\label{gdef}
g_z=-\frac{\lambda|t_z|}{2\omega_z}
\end{equation}
describes the absorption or emission of a resonator photon  accompanying each inter-dot tunneling process. Below,
the rates $\kappa_{\rm in,out}$ refer to photon leakage from the resonator into the microwave input or output lines, with the total leakage rate
\begin{equation}\label{totalkappa}
    \kappa=\kappa_{\rm in}+\kappa_{\rm out}.
\end{equation}
Adding a single-tone input drive with frequency $\omega$ and intensity $I_0$ incident on the resonator,
\begin{equation}\label{drive}
H_{\rm dr}=i\sqrt{\kappa_{\rm in}I_0}\left [ e^{-i\omega t} a^\dagger- e^{i\omega t} a \right ],
\end{equation}
and transforming to the rotating frame of the drive, one arrives at the full strong-coupling Hamiltonian
\begin{equation}\label{scH}
H_{\rm sc} = \frac{\Delta_z}{2} \tau_z+\Delta_0 a^\dagger a + g_z 
(a\tau_++a^\dagger \tau_-)+i\sqrt{\kappa_{\rm in}I_0}(a^\dagger-a),
\end{equation}
with the shifted dot and resonator frequencies
\begin{equation}\label{Deltadef}
    \Delta_z=2\omega_z-\omega,\quad \Delta_0=\omega_0-\omega.
\end{equation}  
Taking into account dephasing and relaxation for the hybridized double dot as well as photon leakage effects
within the standard Lindblad master equation approach \cite{Breuer2007},
the dynamics of the density matrix $\rho(t)$ describing the coupled system defined by the double dot and the resonator 
obeys the Lindblad equation
\begin{eqnarray}\nonumber
\dot\rho&=&-i[H_{\rm sc},\rho] + \frac{\Gamma_\phi}{2}{\cal D}[\tau_z]\rho+ \\ \label{lindblad}
&+& \gamma \left\{(n_{\rm th}+1) {\cal D}[\tau_-]\rho + n_{\rm th} 
{\cal D}[\tau_+]\rho\right\} +\\\nonumber &+& \kappa  
\left \{(N_{\rm th}+1){\cal D}[a]\rho 
+ N_{\rm th} {\cal D}[a^\dagger] \rho\right\},
\end{eqnarray}
where ${\cal D}[L]\rho=L\rho L^\dagger-\frac12\{L^\dagger L,\rho\}$ is the Lindblad dissipator. The terms $\sim \kappa$ model photon leakage, and the rates $\Gamma_\phi$ and $\gamma$ quantify  dephasing and relaxation effects on the double dot, respectively. 
Assuming that a uniform temperature $T$ is established by the   environment, the thermal occupation factors in Eq.~\eqref{lindblad} are given by
\begin{equation}
    n_{\rm th}=\frac{1}{e^{2\omega_z/T}-1},\quad N_{\rm th}=
\frac{1}{e^{\omega_0/T}-1}.
\end{equation}
From Eq.~\eqref{lindblad}, the expectation values $\langle a(t)\rangle={\rm Tr}[a\rho(t)]$ and $\langle \tau_-(t)\rangle$ satisfy the coupled equations of motion
\begin{eqnarray}\nonumber
    \langle \dot a\rangle &=&-i\Delta_0\langle a\rangle-ig_z\langle\tau_-\rangle
    +\sqrt{\kappa_{\rm in} I_0} - \frac{\kappa}{2} \langle a\rangle,\\ \label{eom}
    \langle \dot \tau_-\rangle &=&- i\Delta_z\langle\tau_-\rangle+ ig_z\langle a\tau_z\rangle -  \Gamma_{\rm tot}\langle \tau_-\rangle,
\end{eqnarray}
with the total decay rate 
\begin{equation}\label{rattol}
    \Gamma_{\rm tot}=\Gamma_\phi+\frac{\tilde\gamma}{2},\quad \tilde\gamma= (2n_{\rm th}+1)\gamma.
\end{equation}
However, due to the presence of $\langle a\tau_z \rangle$ in the equation for $\langle\dot\tau_-\rangle$, Eqs.~\eqref{eom} couple to higher-order hierarchies 
of expectation values and an exact solution appears to be out of reach.    

In order to make approximate but analytical progress for the steady-state (ss) solutions of Eq.~\eqref{eom}, reached at long times with $\langle \dot a\rangle_{\rm ss}=\langle\dot \tau_-\rangle_{\rm ss}=0$, Ref.~\cite{Plugge_2017} has truncated the hierarchy 
by applying a semiclassical decoupling approximation in Eq.~\eqref{eom},
\begin{equation}\label{approx}
    \langle a\tau_z\rangle_{\rm ss} \approx \langle a\rangle_{\rm ss} \langle \tau_z\rangle_{\rm ss},
\end{equation}
where  the decoherence effects in Eq.~\eqref{lindblad} imply
\begin{equation}
    \langle \tau_z\rangle_{\rm ss}=-\tanh(\omega_z/T).
\end{equation}
From Eqs.~\eqref{eom} and \eqref{approx}, the steady-state expectation values are then given by
\begin{eqnarray} \nonumber
\langle a\rangle_{\rm ss} &=& \frac{-i\sqrt{\kappa_{\rm in}I_0}}{-\frac{i}{2}\kappa+\Delta_0-\chi_z \langle\tau_z\rangle_{\rm ss}},\\  \label{ssvalues}
\langle \tau_- \rangle_{\rm ss} &=& -\frac{\chi_z}{g_z} \langle a\rangle_{\rm ss}
\langle\tau_z\rangle_{\rm ss},
\end{eqnarray}
where the parity-dependent charge susceptibility is 
\begin{equation}\label{susc}
\chi_z (\omega) = -\frac{g_z^2}{\Delta_z-i\Gamma_{\rm tot}},
\end{equation}
with $g_z$ in Eq.~\eqref{gdef} and $\Delta_z$ in Eq.~\eqref{Deltadef}.
Under the approximation \eqref{approx}, 
the steady-state transmission amplitude in the output line of Fig.~\ref{fig1}(a) thus follows as \cite{Plugge_2017}
\begin{equation}\label{strongcouplingA}
    A^{({\rm sc})}_\omega= \frac{\sqrt{\kappa_{\rm out}}\langle a\rangle_{\rm ss}}{\sqrt{I_0}}
    = \frac{-i\sqrt{\kappa_{\rm in}\kappa_{\rm out}}}{\Delta_0-\frac{i}{2}\kappa-\chi_z \langle\tau_z\rangle_{\rm ss}},
\end{equation}
where ``sc'' emphasizes that this result applies to the strong-coupling limit. 
The susceptibility $\chi_z$ captures the parity-dependent reactive response arising from charge fluctuations, while $A_\omega$ characterizes the parity-dependent two-port response to a microwave drive. 
Parity readout of $\hat z$ is then possible either by tracking the peak in $|A_\omega|^2$
at the $z$-dependent frequency $\omega=\Omega_z$, where $\Omega_z$ follows from Eq.~\eqref{strongcouplingA} by minimizing 
$|\omega_0-\omega- \langle \tau_z\rangle_{\rm ss} {\rm Re}(\chi_z)|$, or by 
measuring the $z$-dependent phase shift $\delta\phi_z=-{\rm arg}(A_\omega)$ of the outgoing signal.
The strong-coupling regime allows for fast high-contrast parity readout but requires sizable coupling $\lambda$ and accurate experimental control over the tunable parameters of the device.  Readout of the MBQ parity $z=\pm$ is thus possible by exploiting the $z$-dependence of $\chi_z$.

It is one of the   goals of the present work to check the accuracy of the semiclassical 
approximation \eqref{approx}.  We can thereby judge the accuracy of expressions like Eq.~\eqref{strongcouplingA}, which allow one to interpret experimental data for 
parity readout in the strong-coupling regime.  

\subsubsection{Off-resonant conditions}

Next we turn to the off-resonant case, realizing the dispersive regime of MBQ parity readout, see also Refs.~\cite{Ohm2015,Yavilberg2015,Plugge_2017,Karzig2017}. Recent experimental readout schemes \cite{Aghaee2025,vanloo2025singleshotparityreadoutminimal} have employed this regime.  The dispersive regime is realized for $\lambda\ll {\rm min}(\omega_0,|\omega_0\pm 2\omega_z|)$, so that only virtual photon processes are important. Expanding $H$ in Eq.~\eqref{hamfull} up to second order in $\{\lambda/\omega_0,\lambda/\omega_{z}\}$, one finds the effective coupled dot-resonator Hamiltonian (up to a $z$-dependent constant) \cite{Plugge_2017}
\begin{equation}\label{dispham}
    H_{\rm eff}= \left(\omega_z+\frac{\delta_z+\chi_z}{2}
    \right)\tau_z + (\omega_0+\chi_z\tau_z)a^\dagger a ,
\end{equation}
where $\chi_z(\omega)$ now has the $\omega$-independent (static) form
\begin{equation}\label{susc2}
\chi_z =- \frac{g_z^2}{\omega_z-\omega_0^2/4\omega_z}.
\end{equation} 
We also use $\delta_z=\lambda^2\varepsilon/(\omega_0\omega_z)$.
In analogy to Eq.~\eqref{scH}, using $H_{\rm dr}$ in  Eq.~\eqref{drive} and switching to the rotating frame of the drive, we arrive at the full weak-coupling Hamiltonian 
\begin{equation}
    H_{\rm wc}=  
    \frac{\Delta_z+\delta_z+\chi_z}{2}\tau_z+ (\Delta_0+\chi_z \tau_z) a^\dagger a +i\sqrt{\kappa_{\rm in}I_0}(a^\dagger-a).
\end{equation} 
We now again employ the Lindblad equation \eqref{lindblad} with  $H_{\rm sc}\to H_{\rm wc}$ to include dephasing and relaxation effects on the double dot as well as photon leakage.
Instead of Eq.~\eqref{eom}, we arrive at the weak-coupling form of the equation of motion for $\langle a\rangle$,
 \begin{equation} \label{eoma}
 \langle \dot a\rangle =-i\Delta_0\langle a\rangle- i\chi_z\langle a \tau_z \rangle
    +\sqrt{\kappa_{\rm in} I_0} - \frac{\kappa}{2} \langle a\rangle.
\end{equation}
Here, $\langle a\tau_z\rangle$ satisfies a closed equation of motion,
\begin{eqnarray}\nonumber
 \frac{d}{dt} \langle a \tau_z\rangle  &=& - \left(i\Delta_0+\frac{\kappa}{2} +\tilde\gamma\right)\langle a\tau_z \rangle - \left(i\chi_z +\gamma\right)\langle a\rangle \\
    &+& \sqrt{\kappa_{\rm in}I_0} \langle \tau_z\rangle,\label{eomatz}
\end{eqnarray}
with $\tilde\gamma$ in Eq.~\eqref{rattol}. 
If one again applies the semiclassical approximation \eqref{approx}, the result for $A_\omega$ in the weak-coupling (wc) regime corresponding to dispersive (off-resonant) MBQ readout follows directly from Eq.~\eqref{eoma},
\begin{equation}\label{weakcouplingA}
    A_\omega^{{\rm (wc)}}
    = \frac{-i\sqrt{\kappa_{\rm in}\kappa_{\rm out}}}{\Delta_0-\frac{i}{2}\kappa+ \chi_z \langle\tau_z\rangle_{\rm ss}},
\end{equation}
which formally coincides with the strong-coupling expression \eqref{strongcouplingA} apart from a sign change in the $\chi_z$ term.  Parity readout is again possible by 
determining $\chi_z$ from measurements of $A_\omega$.

However, since Eq.~\eqref{eomatz} allows one to compute $\langle a\tau_z\rangle_{\rm ss}$ without further truncation,  the exact form of $A_\omega=A_{\omega}^{({\rm wc, ex.})}$ 
can be determined analytically. With $\tilde \gamma$ in Eq.~\eqref{rattol}, we find
\begin{equation} \label{weakcouplingexactA}
    A_{\omega}^{({\rm wc, ex.})}
    = \frac{-i\sqrt{\kappa_{\mathrm{in}}\kappa_{\mathrm{out}}}
    \left[\Delta_0 -i(\frac{\kappa}{2}
    + \tilde\gamma)-\chi_z\langle \tau_z\rangle_{\mathrm{ss}} \right]}
    {\left(\Delta_0 -\frac{i}{2}\kappa\right)
    \left[\Delta_0 -i\left(\frac{\kappa}{2} + \tilde\gamma\right)\right]-(\chi_z - i\gamma)\chi_z
    }.
\end{equation}
From this expression, one recovers the approximate result \eqref{weakcouplingA} (obtained under the semiclassical approximation) as follows. We first assume that relaxation effects on the double dot are subleading, such that terms $\propto \gamma,\tilde\gamma$ in Eq.~\eqref{weakcouplingexactA} can be neglected.  Using $|\chi_z \langle \tau_z\rangle_{\rm ss}| \ll |\Delta_0-i\frac{\kappa}{2}|$, which is expected to hold in the weak-coupling regime, see Eq.~\eqref{susc2}, we may write $1-\frac{\chi_z\langle\tau_z\rangle_{\rm ss}}{\Delta_0-i\frac{\kappa}{2}}\simeq \left( 1+\frac{\chi_z\langle\tau_z\rangle_{\rm ss}}{\Delta_0-i\frac{\kappa}{2}}\right)^{-1}$.  In addition, we neglect the term $\propto (\chi_z-i\gamma)\chi_z$ 
in the denominator of Eq.~\eqref{weakcouplingexactA} in view of the smallness of $\chi_z$. 
With these approximations, Eq.~\eqref{weakcouplingexactA} reduces to Eq.~\eqref{weakcouplingA}.  We conclude that the approximation \eqref{approx} is justified for very weak relaxation effects on the dot and for very small susceptibility $\chi_z$.  For a more detailed comparison of 
Eqs.~\eqref{weakcouplingA} and \eqref{weakcouplingexactA}, see Sec.~\ref{sec3b}.

\subsection{Quantum capacitance}\label{sec2c}

We now turn to the setup in Fig.~\ref{fig1}(b), where MBQ parity readout is performed by measuring the $z$-dependent quantum capacitance $C_{Q;z}$.  If the tunneling amplitude $t_0$ through the reference arm is much bigger than the cotunneling amplitude $t_1$, the two quantum dots are effectively hybridized to form a single dot which is tunnel-coupled to both Majorana operators $\gamma_2$ and $\gamma_3$.  The energy shift of this dot then depends on $\hat z$ which can be read out by measuring $C_{Q;z}$.  
Interestingly, our results obtained for the susceptibility $\chi_z$, which determines the charge reflectometry parity readout scheme in Sec.~\ref{sec2b}, can be translated into  results for the effective parity-dependent quantum capacitance $C_{Q;z}$.  Similarly, the two-port transmission amplitude $A_\omega$ can be recast as one-port reflection coefficient $S^M_{11}$, corresponding to the measured signal in the one-port reflectometry setup in Ref.~\cite{vanloo2025singleshotparityreadoutminimal}, see Sec.~\ref{sec2d}.

The Hamiltonian in Eq.~\eqref{hamfull} describes the double dot coupled to the resonator in the presence of the parity-dependent hopping amplitude $t_z$ both in the weak-coupling and the strong-coupling regime, including the crossover between both limits.  The MBQ parity is 
here contained in the term $H_I=g_z (a+a^\dagger) \tau_x$.  For the quantum capacitance readout sketched in Fig.~\ref{fig1}(b), the corresponding form of $H_I$, denoted by $\tilde H_I$ below, is obtained
by transforming the capactive coupling between the resonator and dot 2 into the Rabi basis of the double dot. We then find 
\begin{equation}\label{tildeHi}
    \tilde H_I= -\alpha e \sqrt{\frac{\omega_0}{2C_z}} (a+a^\dagger) \tau_x,
\end{equation} 
where $C_z$ is the total effective capacitance of the readout circuit, which includes the resonator capacitance $C_r$ and the quantum capacitance $C_{Q;z}$. The quantity $\alpha$ in Eq.~\eqref{tildeHi} is a lever arm \cite{Aghaee2025}.  By comparing $H_I$ and $\tilde H_I$, we obtain the correspondence
\begin{equation}\label{gqc}
g_z =-\alpha e \sqrt{\frac{\omega_0}{2C_z}}.
\end{equation}
We now consider the linear-response regime, where the 
induced charge on dot 2 is connected to the applied drive voltage $V(\omega)$ through the charge susceptibility $\tilde\chi_z(\omega)$ of the double dot.   
Following the unified framework of Ref.~\cite{Peri_2024}, the admittance is then given by
\begin{equation}
    Y(\omega) = -i\omega(\alpha e)^2\tilde\chi_z,
    \label{eq:Y_peri}
\end{equation}
which identifies the reactive part of the response, $Y(\omega)=i\omega C_{Q;z}$, as an effective $z$-dependent quantum capacitance, 
\begin{equation}
    C_{Q;z}= -(\alpha e)^2 \tilde\chi_{z}.
    \label{eq:CQ_from_chinn}
\end{equation}
In the device circuitry of Fig.~\ref{fig1}(b),
the reactive response due to the double dot provides a small  parity-dependent quantum capacitance contribution, $C_{Q;z}$, in parallel with the resonator capacitance $C_r$. We note that
the resonator susceptibility $\chi_z$ is related to the susceptibility of the double dot by the relation $\chi_z=g_z^2\tilde \chi_z$.
With $C_{Q;z}\ll C_r$, from Eq.~\eqref{eq:CQ_from_chinn} we arrive at the relation, see also Ref.~\cite{Aghaee2025},
\begin{equation}\label{eq:CQ_final_prefactor}
C_{Q;z}(\omega)= -\frac{2C_r}{\omega_0}\chi_z(\omega) .
\end{equation}
The minus sign implies that a positive reactive susceptibility
(${\rm Re}(\chi_z)>0$) corresponds to a reduction of the effective quantum capacitance, implying an upward shift of the resonator frequency.  By virtue of Eq.~\eqref{eq:CQ_final_prefactor},
 results for the charge reflectometry setup are directly connected to quantum capacitance readout protocols.  In both approaches, $\chi_z$ is the central $z$-dependent quantity of interest.

\subsection{Single-port reflectometry}\label{sec2d}

We here briefly show that the parity-dependent single-port reflectometry signal $S_{11}^{M}$ measured in Ref.~\cite{vanloo2025singleshotparityreadoutminimal} is also equivalent to
the knowledge of $A_\omega$. We start from the standard input-output description of a resonator connected to two ports as in Fig.~\ref{fig1}(a), where the cavity field $a(\omega)$ in frequency space is connected to the incoming and outgoing fields $b_{\rm in,out}(\omega)$ and $c_{\rm in,out}(\omega)$ in the two ports  through the relations   \cite{Clerk_2010}
\begin{eqnarray}
a(\omega) &=& \mathcal{R}(\omega)\!\left[
   \sqrt{\kappa_{\mathrm{in}}}b_{\mathrm{in}}(\omega)
 + \sqrt{\kappa_{\mathrm{out}}}c_{\mathrm{in}}(\omega)
 \right],\nonumber \\ 
\label{eq:a_general}
b_{\mathrm{out}}(\omega) &=& b_{\mathrm{in}}(\omega)
 - \sqrt{\kappa_{\mathrm{in}}}a(\omega),\\ \nonumber
c_{\mathrm{out}}(\omega) &=& c_{\mathrm{in}}(\omega)
 - \sqrt{\kappa_{\mathrm{out}}}a(\omega).
\end{eqnarray}
The complex function $\mathcal{R}(\omega)$ characterizes the internal resonator response and contains the parity-dependent signal.
In this two-port configuration, the drive is applied at the input port, i.e., $c_{\mathrm{in}}=0$. The transmitted amplitude in the output port is then given by
\begin{equation}
S_{21}(\omega) = \frac{\langle c_{\mathrm{out}}(\omega)\rangle}
        {\langle b_{\mathrm{in}}(\omega)\rangle}
 = -\sqrt{\kappa_{\mathrm{in}}\kappa_{\mathrm{out}}}\mathcal{R}(\omega).
\label{eq:S21_basic}
\end{equation}
For drive amplitude $\langle b_{\mathrm{in}}(\omega)\rangle=\sqrt{I_0}$,
the steady-state cavity field is then given by
$\langle a(\omega)\rangle_{\mathrm{ss}}
   = \sqrt{\kappa_{\mathrm{in}}I_0}\mathcal{R}(\omega),$
and we arrive at
\begin{equation}
A_\omega  = \sqrt{\frac{\kappa_{\mathrm{out}}}{I_0}}
     \langle a(\omega)\rangle_{\mathrm{ss}}
    = S_{21}(\omega).
\end{equation}

Similarly, for a single-port configuration, the reflected amplitude at the port follows as
\begin{equation}
S_{11}(\omega)   = \frac{\langle b_{\mathrm{out}}(\omega)\rangle}
          {\langle b_{\mathrm{in}}(\omega)\rangle}
   = 1 - \kappa_{\mathrm{in}}\mathcal{R}(\omega).
\label{eq:S11_general}
\end{equation}
Using Eqs.~\eqref{eq:S21_basic} and \eqref{eq:S11_general}, we arrive at  the general  relation
\begin{equation}
S_{11}(\omega)
   = 1 + \sqrt{\frac{\kappa_{\mathrm{in}}}{\kappa_{\mathrm{out}}}}
     S_{21}(\omega)
   = 1 + \sqrt{\frac{\kappa_{\mathrm{in}}}{\kappa_{\mathrm{out}}}}
     A_\omega .
\label{eq:S11_vs_Aw}
\end{equation}
We emphasize that $S_{11} = S_{11}^{M}$ contains again the  
parity-dependent reflectometry signal. This readout scheme was used in Ref.~\cite{vanloo2025singleshotparityreadoutminimal}.

\section{Results}\label{sec3}

In this section, we first derive an expression for the susceptibility $\chi_z$
valid in the full crossover regime from weak to strong coupling, see Sec.~\ref{sec3a}.  This expression is of particular interest for  quantum capacitance readout. We compare it to the corresponding
approximate expressions valid in the weak- and strong-coupling limits, see Eqs.~\eqref{susc2} and \eqref{susc}, respectively, 
in order to estimate their accuracy in different parameter regimes.  In Sec.~\ref{sec3b}, 
we then turn to an assessment of the accuracy of expressions based on the semiclassical 
approximation \eqref{approx}, which has been employed for deriving the charge reflectometry amplitude $A_\omega$ in Sec.~\ref{sec2}.  
Specifically, we compare the strong-coupling
and weak-coupling expressions for $A_\omega$ in Eqs.~\eqref{strongcouplingA}
and \eqref{weakcouplingA} to exact results.  In the strong-coupling limit, the exact steady state is obtained by numerical simulation of the full Lindblad equation
\eqref{lindblad}. In the weak-coupling limit, we instead compare to the exact result for $A_\omega$ in Eq.~\eqref{weakcouplingexactA}. 

\subsection{Susceptibility for the full crossover}\label{sec3a}

We start by computing the dynamical susceptibility associated with the cavity-qubit coupling operator in Eq.~\eqref{hamfull} for the entire crossover from weak to strong coupling. From the linear coupling term $(a+a^\dagger) \hat{D}$ in Eq.~\eqref{hamfull}, we identify
the operator
\begin{equation}\label{hatD}
    \hat{D} = \frac{\lambda}{2}\left(1-\frac{\varepsilon}{\omega_z}\tau_z -\frac{|t_z|}{\omega_z} \tau_x
    \right).
\end{equation}
We consider this coupling as weak perturbation and treat it by linear response theory.
The susceptibility thus follows from a Kubo formula~\cite{Bruus2004},
\begin{equation}\label{Kubo}
    \chi_z^{\rm full}(t) = -i\,\Theta(t)\,\langle[\hat{D}(t), \hat{D}(0)]\rangle_0 ,
\end{equation}
where the average is taken with respect to the $\lambda=0$ problem.
Since $[\tau_z(t),\tau_z(0)]=0$ in the absence of the coupling, only the term $\propto\tau_x$ in Eq.~\eqref{hatD} contributes to Eq.~\eqref{Kubo}.
We can therefore effectively replace $\hat{D} \rightarrow g_z \tau_x$  with $g_z$ in Eq.~\eqref{gdef} when computing $\chi_{z}^{\rm full}(t)$.  
The correlator in Eq.~\eqref{Kubo} can be analytically evaluated, 
taking into account dephasing and decoherence processes in the Lindblad
equation \eqref{lindblad} via the rate $\Gamma_{\rm tot}$ in Eq.~\eqref{rattol}. After a Fourier transform, we find
\begin{equation}\label{suscgen}
    \chi^{\mathrm{full}}_z(\omega)  = g_z^2 \left[
    \frac{1}{i\Gamma_{\mathrm{tot}} - (2\omega_z - \omega)}    -
    \frac{1}{i\Gamma_{\mathrm{tot}} + (2\omega_z + \omega)}
    \right] .
\end{equation}
This expression describes the susceptibility for the full crossover from weak to strong coupling within linear-response theory, i.e., taking into account terms up to the order $g_z^2$.
The first term in Eq.~\eqref{suscgen} corresponds to the resonant contribution near $\omega \simeq 2\omega_z$,
which is kept within RWA for the strong-coupling case, while the second term represents the counter-rotating contribution.
In the near-resonant limit, $\omega \approx  2 \omega_z$, the off-resonant second term can be neglected, leading to the susceptibility expression in Eq.~\eqref{susc} for the strong-coupling limit.
Conversely, setting $\omega\approx \omega_0$ in Eq.~\eqref{suscgen} and expanding both denominators for large detuning, $|\omega_0\pm 2\omega_z|\gg {\rm max}\{|g_z|,\Gamma_{\mathrm{tot}}\}$, 
we recover the leading-order static dispersive pull in the weak-coupling regime, see Eq.~\eqref{susc2}.

\begin{figure}
    \includegraphics[width=0.5\textwidth]{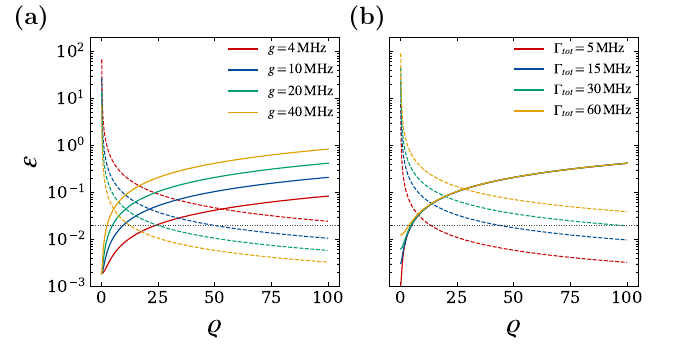}
    \caption{Relative deviation $\varepsilon$ in Eq.~\eqref{vareps} vs normalized detuning 
    $\varrho$ in Eq.~\eqref{detuning}, comparing the full susceptibility expression \eqref{suscgen} to approximate expressions. Note the logarithmic scales for $\varepsilon$.
    Here  $\varrho$ is changed by ramping up $\omega_0-2\omega_z$ at given coupling strength $g$.
    Since $\varepsilon(-\varrho)\simeq \varepsilon(\varrho)$, we depict only $\varepsilon(\varrho\ge 0)$. 
    Solid curves show $\varepsilon$ using the strong-coupling approximation for $\chi_z$ in Eq.~\eqref{susc}, and dashed curves refer to the weak-coupling approximation in Eq.~\eqref{susc2}.
    The horizontal dotted line indicates a $2\%$ error threshold.  
    (a) $\varepsilon$ vs $\varrho$ for several 
    values of $g$ in the range $4~{\rm MHz}\le g\le 40$~MHz with fixed total decoherence rate $\Gamma_{\rm tot}=9$~MHz.
    (b) Same as (a) but for several values of $\Gamma_{\mathrm{tot}}$ in the range $9~{\rm MHz}\le \Gamma_{\rm tot}\le 60$~MHz with fixed coupling strength $g=20$~MHz. 
    }  \label{fig2}
\end{figure}

To quantify the validity range of the approximate susceptibility expressions in Eqs.~\eqref{susc} and \eqref{susc2}, let us compare them to the full susceptibility \eqref{suscgen} in more detail.  
We quantify the accuracy of each approximation by evaluating the relative deviation, 
\begin{equation}\label{vareps}
    \varepsilon = \frac{|\chi_z^{\mathrm{full}}-\chi_z^{\mathrm{appr}}|}{|\chi_z^{\mathrm{full}}|} ,
\end{equation}
computed separately for the strong-coupling limit, with $\chi_z^{\mathrm{appr}}$ in Eq.~\eqref{susc}, and for the dispersive limit, see Eq.~\eqref{susc2}.
Keeping the $z$-dependence implicit, i.e., writing the coupling strength as $g=g_z$, 
and for simplicity setting $\omega=\omega_0$,
the detuning is quantified by the dimensionless parameter
\begin{equation}\label{detuning}
    \varrho=\frac{\omega_0- 2\omega_z}{g},
\end{equation} 
where $|\varrho|\ll 1$ corresponds to the resonant regime and $|\varrho|\gg 1$ 
characterizes the weak-coupling limit. Figure~\ref{fig2} tests the accuracy of Eqs.~\eqref{susc} and \eqref{susc2}, which apply in the strong- and weak-coupling regime, respectively, across the full crossover as function of $\varrho$. 
Typical experimental values for $\omega_0$ and $\omega_z$ are in the GHz regime. The decoherence rate $\Gamma_{\rm tot}$ and the coupling strength $g$ are taken in the MHz range. Those values are consistent with Majorana box qubit proposals and recent experiments on InAs-Al hybrid devices~\cite{Plugge_2017,Aasen2016,Aghaee2025}.

In Fig.~\ref{fig2}(a), the coupling strength $g$ is varied for a fixed value of $\Gamma_{\mathrm{tot}}$. 
Close to resonance ($\varrho\simeq 0$), the strong-coupling limit is (as expected) accurately captured by Eq.~\eqref{susc} 
since the counter-rotating contribution in Eq.~\eqref{suscgen} is far off resonance.
At the same time, the dispersive result \eqref{susc2} breaks down. 
As $\varrho$ increases, the weak-coupling expression \eqref{susc2} becomes progressively more accurate, while the strong-coupling result \eqref{susc} becomes increasingly less accurate.
Decreasing the coupling strength $g$ broadens the interval in $\varrho$ over which neither approximation is accurate, indicating that the width of the crossover region depends on $g$.
For instance, for $g=10$~MHz, the error $\varepsilon$ exceeds an error threshold
of $2\%$ for both approximations within the detuning window $10\alt \varrho \alt 50$, while for $g=40$~MHz, this window shrinks to $2\alt \varrho\alt 12$.

Next, Fig.~\ref{fig2}(b) illustrates the effect of the decoherence rate $\Gamma_{\rm tot}$ at fixed coupling. 
Increasing $\Gamma_{\mathrm{tot}}$ broadens the resonance in Eq.~\eqref{suscgen}, thereby softening the breakdown of the weak-coupling result for small $\varrho$ and generally increasing the error level for both approximations.
Strong decoherence rates smoothen out the resonant structure and tend to invalidate both 
approximations.  For MBQ parity readout, one thus has to work with small  $\Gamma_{\rm tot}$.

To summarize, Fig.~\ref{fig2} shows that Eq.~\eqref{suscgen} reproduces the expected behavior in both limits. In the near-resonant regime, it reduces to the strong-coupling susceptibility \eqref{susc}, while for large detuning, it approaches the weak-coupling result \eqref{susc2}.
The full susceptibility expression \eqref{suscgen} provides a consistent description across the full parameter range.

\subsection{Tests of the semiclassical approximation}\label{sec3b}

Next we turn to tests of the validity of the semiclassical approximation \eqref{approx}.  In order to do so, we compare the approximate steady-state predictions in Sec.~\ref{sec2b} to
exact reference values.  In the strong-coupling limit, these are obtained from a numerical  solution for the steady state of the Lindblad equation \eqref{lindblad}, retaining the full coupled photon-qubit dynamics. In the weak-coupling limit, we instead use the exact results 
discussed in Sec.~\ref{sec2b}.
The comparison is performed both for photon-qubit operator correlations encoded by
$\langle a\tau_z\rangle_{\rm ss}$ and for the complex transmission amplitude $A_\omega$ used for MBQ parity readout in charge reflectometry.

First, we recall that the semiclassical approximation \eqref{approx} assumes
$\langle a\tau_z\rangle_{\rm ss} \approx \langle a\rangle_{\rm ss} \langle\tau_z\rangle_{\rm ss}$, thus neglecting connected photon-qubit correlations.
As dimensionless measure for the quality of this approximation, we study the normalized connected correlator
\begin{equation}\label{var1}
    \varepsilon_{a\tau_z}  = \frac{\big|\langle a\tau_z\rangle_{\rm s}-\langle a\rangle_{\rm ss}\langle\tau_z\rangle_{\rm ss}\big|}
    {\left|\langle a \rangle_{\rm ss}\right|},
\end{equation}
which probes the importance of correlations discarded in the approximation. (One could alternatively use  $|\langle a\rangle_{\rm ss}\langle\tau_z\rangle_{\rm ss}|$
in the denominator, but for the parameter regimes considered below, this makes no significant difference.)
Large values of $\varepsilon_{a\tau_z}$ indicate a breakdown of the semiclassical approximation.
Second, to assess the impact of photon-qubit correlations on measurable quantities, we  define a bounded symmetric relative error for  $A_\omega$,
\begin{equation}\label{var2}
    \varepsilon_\mathrm{A}    =
    \frac{|A_\mathrm{semicl}-A_\mathrm{ref}|}{|A_\mathrm{semicl}|+|A_\mathrm{ref}|},
\end{equation}
where $A_\mathrm{semicl}$ is the semiclassical (approximate) result and $A_\mathrm{ref}$ the exact reference solution.  
In addition, since the phase of $A_{\omega}$ often provides the most
sensitive signature in parity readout experiments, we also consider the normalized phase deviation,
\begin{equation}\label{var3}
    \varepsilon_\phi  =  \frac{\big|\arg(A_\mathrm{semicl})-\arg(A_\mathrm{ref})\big|}{\pi} .
\end{equation}
The combined set $\{\varepsilon_{a\tau_z}, \varepsilon_\mathrm{A}, \varepsilon_\phi\}$  allows us to distinguish between small quantitative deviations in the readout signal and a genuine breakdown of the semiclassical approximation in different parameter regimes.
We perform the comparison for parameters that are  realistic for hybrid Majorana devices such as those studied in Refs.~\cite{Aghaee2023,Aghaee2025,vanloo2025singleshotparityreadoutminimal}.

\begin{figure*}
    \centering
    \includegraphics[width=\textwidth]{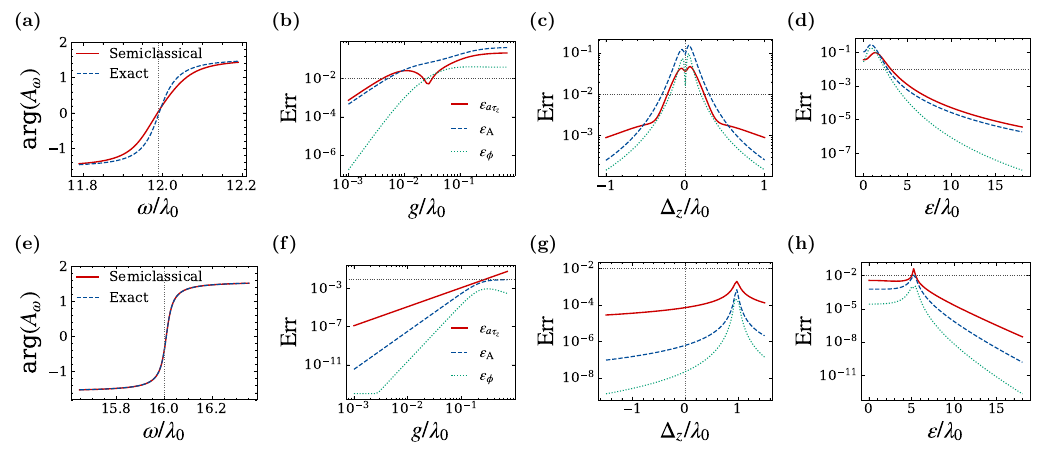}
    \caption{Comparison of semiclassical and exact results in the strong- and weak-coupling limits. We use the dimensionless numbers $\{ \varepsilon_{a\tau_z}, \varepsilon_A, \varepsilon_{\phi}\}$ in Eqs.~\eqref{var1}, \eqref{var2}, and \eqref{var3}, which are shown on logarithmic scales.
    Employing the scale $\lambda_0 = 0.2~\mathrm{GHz}$, we assume the tunnel couplings $|t_0| = 5 \lambda_0$ and $|t_1| = \lambda_0$. For $z=+1$ and $\varphi=0$, we have $|t_z| = |t_0 + t_1|=6\lambda_0$. The cavity linewidth is $\kappa = 0.04 \lambda_0$.  Results are shown at temperature $T=30$~mK.\\
    \textsl{Panels (a)--(d):} \emph{Strong-coupling limit.} Unless noted otherwise, we choose the quantum-dot detuning $\varepsilon=0$, see Eq.~\eqref{hamiltonian}, with $\omega_z$ resulting from Eq.~\eqref{omegaz}, the bare cavity frequency $\omega_0 = 11.99 \lambda_0$,  the rates $\gamma = 0.003 \lambda_0$ and $\Gamma_\phi = 0.1 \lambda_0$, the drive frequency $\omega=\omega_0$, and the cavity-dot coupling $\lambda = 0.1 \lambda_0$, with $g$ obtained from Eq.~\eqref{gdef}.
    Panel (a) shows the phase of $A_\omega$ vs $\omega$. The dotted vertical line indicates the resonance at $\omega=\omega_0$.
    Panels (b), (c), and (d) show the error measures as function of $g$, of the detuning $\Delta_z=2\omega_z-\omega$, and of $\varepsilon$, respectively. 
    Dottel horizontal lines mark an error level of $1\%$.  \\
    \textsl{Panels (e)--(h):} Same as (a)--(d) but for the \emph{weak-coupling limit.} 
    Unless noted otherwise, we here put $\varepsilon = 6 \lambda_0$,  $\lambda = 0.25 \lambda_0$, $\gamma = 0.015 \lambda_0$, $\Gamma_\phi = 0.05 \lambda_0$, $\omega_0 = 16 \lambda_0$, and $\omega=\omega_0$. In that case, $|\omega_0 - 2\omega_z|\sim {\cal O}(\lambda)$. 
    }  \label{fig3}
\end{figure*}

\subsubsection{Strong-coupling limit}

In the strong-coupling limit, the three error measures introduced above are evaluated by comparing the semiclassical predictions in Sec.~\ref{sec2b} to the numerically exact solution for the steady state of the Lindblad equation.  Our results are shown in Fig.~\ref{fig3}(a)--(d) for a specific parameter set, where the system is in the  strong-coupling regime such that coherent exchange dominates over dissipative broadening \cite{Blais2004}. Figure~\ref{fig3}(a) compares semiclassical and exact results for the transmission phase $\arg(A_\omega)$ as function of the drive frequency $\omega$.
The semiclassical approximation captures the characteristic phase rotation, but close to resonance, $\omega\simeq \omega_0$, quantitative deviations are visible.
In particular, the slope of the phase response differs, 
indicating that coherent photon-qubit hybridization effects modify 
the dynamical response beyond semiclassics. Nonetheless, the deviations remain of moderate size. In particular, the phase error $\varepsilon_\phi$ remains below $10 \%$ across the shown frequency range. 

A systematic comparison between the semiclassical and the exact solution is provided in Fig.~\ref{fig3}(b)--(d), where we show the dependence of the three error measures [see Eqs.~\eqref{var1}, \eqref{var2}, and \eqref{var3}] on the coupling strength $g$, on the qubit-drive detuning $\Delta_z$ in Eq.~\eqref{Deltadef}, and on the double-dot detuning $\varepsilon$ in Eq.~\eqref{hamiltonian}, respectively.  In particular, Fig.~\ref{fig3}(b) shows that for small $g$,
all error measures grow with increasing $g$, indicating the growing importance 
of coherent mixing and correlation effects.
Interestingly, $\varepsilon_{a\tau_z}$ exhibits a minimum at an intermediate value of $g$.
This dip signals a crossover regime, where dissipative broadening suppresses the buildup of photon-qubit correlations while coherent hybridization does not yet dominate.
However, in view of Eq.~\eqref{var1}, this minimum can also be partially influenced by the fact that $|\langle a\rangle_{\rm ss}|$ varies with $g$.
For larger values, $g \gtrsim (\kappa, \gamma)$, hybridization effects dominate and all 
error measures grow with increasing $g$.  For the entire parameter range in Fig.~\ref{fig3}(b), all error measures remain below the $10\%$ margin.

Next, in Fig.~\ref{fig3}(c), we show the dependence of the error measures on the detuning $\Delta_z=2\omega_z-\omega$.
All error measures exhibit maxima near $\Delta_z = 0$,
where coherent exchange between photons and the double dot becomes most efficient, leading to maximal values of the connected photon-qubit correlations, $\big|\langle a\tau_z\rangle_{\rm ss}-\langle a\rangle_{\rm ss}\langle\tau_z\rangle_{\rm ss}\big|$ and, consequently, to shortcomings of Eq.~\eqref{approx}.
We note that a small antidip is visible in the error measures at $\Delta_z=0$. 
This feature can be rationalized by noting that $\chi_z$ becomes purely imaginary for $\Delta_z=0$, i.e., the double dot causes only additional damping but no frequency shift of the cavity.  The semiclassical approximation is then slightly better, resulting in a smaller
error. We note that even close to the peaks centered near $\Delta_z=0$, all three errors remain again
below the $10\%$ level. With increasing $|\Delta_z|$, they rapidly drop below $1\%$.

Finally, Fig.~\ref{fig3}(d) shows the dependence on the scale $\varepsilon$, which enters the Rabi frequency $\omega_z$ in Eq.~\eqref{omegaz}.  Since the results for $\varepsilon\to -\varepsilon$ are identical, we consider only $\varepsilon\ge 0$ in Fig.~\ref{fig3}(d).
For small $\varepsilon$, where the effective qubit splitting is minimal and hybridization effects are strongest, all error measures reach their maximal values.
Increasing $\varepsilon$ enhances the energy separation between the qubit and the cavity, thereby suppressing coherent exchange processes.
Correspondingly, all error measures decrease rapidly with increasing $\varepsilon$.

 For the considered parameter values, the errors therefore always remain in the few-percent range, never exceeding the $10\%$ level.
While the semiclassical approximation is not exact, it remains rather accurate and the corresponding expressions can be 
used for interpreting MBQ parity readout experiments unless one works in a  low-contrast regime where $z=\pm 1$ outcomes are hard to distinguish.
In such cases, one should employ a numerical solution of the full Lindblad equation to compare experimental data to theoretical predictions.
We emphasize that deviations from semiclassical estimates become larger near resonance and/or for large coupling strengths.

\subsubsection{Weak-coupling regime}

We now turn to the dispersive weak-coupling regime, 
see Fig.~\ref{fig3}(e)--(h), where the cavity and the double dot are energetically detuned and coherent hybridization effects are suppressed. In contrast to the strong-coupling limit, the semiclassical approximation can now be benchmarked against analytical expressions, see Sec.~\ref{sec2b}.
We again assess the accuracy of the approximation by means of the error measures $\{\varepsilon_{a\tau_z},\varepsilon_{\mathrm A},\varepsilon_\phi\}$. For the parameters in Fig.~\ref{fig3}(e)--(h), we have $ g \ll |\omega_0 - 2\omega_z|$ and 
$g \lesssim ( \kappa, \gamma)$. In this weak-coupling regime, the semiclassical approach effectively corresponds to retaining the leading 
dispersive frequency shift induced by the double dot on the cavity   while higher-order dynamical corrections are suppressed~\cite{Blais2004,Smith2020}.

In Fig.~\ref{fig3}(e), both the approximate and the exact result for $\arg(A_\omega)$ are shown as function of the drive frequency $\omega$.
In contrast to the strong-coupling limit in Fig.~\ref{fig3}(a), the phase response is reproduced almost perfectly by the semiclassical approximation, with no visible discrepancy in slope or magnitude.
In the dispersive limit, the qubit remains close to its equilibrium state and photon-qubit correlations are strongly suppressed.  
As a result, the semiclassical decoupling provides an accurate description of both amplitude and phase.

The error measures are studied in Fig.~\ref{fig3}(f)--(h).
First, in Fig.~\ref{fig3}(f), we study their dependence on the  coupling strength $g$.
All three error measures increase monotonically with the effective coupling strength $g$.
A significant error $\varepsilon_{a\tau_z}$ is visible at weak coupling already, while the amplitude error $\varepsilon_A$ and the phase error $\varepsilon_\phi$ remain extremely small for $g \alt 0.1\lambda_0$ and only approach the $1\%$ threshold for $g \gtrsim 0.2\lambda_0$.
This reflects the fact that the dispersive shift $\chi_z \propto g^2$ sets the scale of the cavity pull.
Increasing $g$ enhances the static frequency shift and increases qubit-cavity correlations not captured by the semiclassical factorization. Because the leading-order dispersive susceptibility is real and frequency-independent, no resonant enhancement occurs and 
all error measures grow monotonically with $g$.
Nonetheless, for almost the full parameter range in Fig.~\ref{fig3}(f), the three error measures, including $\varepsilon_{a\tau_z}$, 
remain  below $1\%$.

In Fig.~\ref{fig3}(g), we show the dependence on the detuning $\Delta_z$.  In contrast to the strong-coupling regime, no pronounced resoance peak occurs near $\Delta_z = 0$.
Instead, all error measures remain orders of magnitude
below the $1\%$ level for the entire range shown.
A smooth maximum is observed around $\Delta_z \simeq \lambda_0$, where the effective detuning between the cavity and the qubit transition becomes minimal within the scanned interval.
At this point, the energy denominator entering the dispersive susceptibility is smallest, such that higher-order dynamical corrections beyond the leading dispersive approximation become pronounced. However, even near this maximum,
the connected correlator $\varepsilon_{a\tau_z}$ and the amplitude error $\varepsilon_{\mathrm A}$ remain at or below the $0.1 \%$ level, with even smaller phase errors $\varepsilon_\phi$.

Similarly, in Fig.~\ref{fig3}(h), the dependence on $\varepsilon$ reveals a pronounced maximum in the error measures around $\varepsilon \simeq 5\lambda_0$.
At this point, the qubit splitting $2\omega_z$ is closest to the cavity frequency within the scanned interval, thereby minimizing the effective detuning $|\omega_0-2\omega_z|$.
For larger $\varepsilon$, the qubit splitting grows and the cavity-qubit detuning increases again.
The dispersive condition is therefore reinforced, leading to rapid suppression of all error measures.
In particular, both the connected correlator $\varepsilon_{a\tau_z}$ and the amplitude error $\varepsilon_{\mathrm A}$ decrease by several orders of magnitude as $\varepsilon$ increases.
Even at the maximum near $\varepsilon \simeq 5\lambda_0$, however, the deviations remain at or below $1\%$.

\begin{figure}
    \centering
    \includegraphics[width=0.4\textwidth]{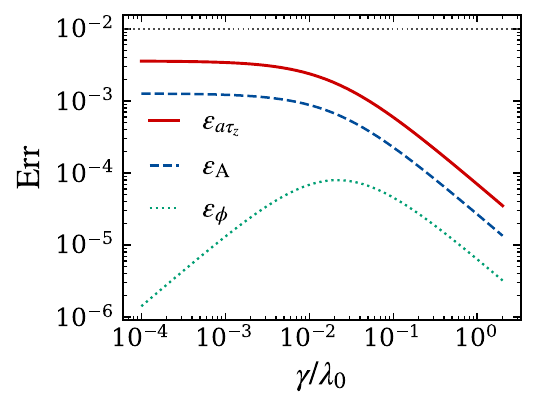}
    \caption{Dependence of the error measures $\varepsilon_{a\tau_z}$, $\varepsilon_{\mathrm A}$, and $\varepsilon_\phi$ on the relaxation rate $\gamma$ of the double dot in the weak-coupling limit.
    Semiclassical results are compared to the exact analytical expressions in Sec.~\ref{sec2b}, see, in particular, Eq.~\eqref{weakcouplingexactA}.
    The dashed horizontal line indicates the $1\%$ level.
    All energies are expressed in units of $\lambda_0 = 0.2~\mathrm{GHz}$.}  \label{fig4}
\end{figure}

In addition, in Fig.~\ref{fig4}, we analyze the dependence of the error measures on the relaxation rate $\gamma$ in order to test the approximation made in deriving the semiclassical expression from Eq.~\eqref{weakcouplingexactA}, where terms proportional to $\gamma$ were dropped.
In the weak-coupling regime considered in Fig.~\ref{fig4}, all error measures remain well below $1\%$. For increasing $\gamma$, we find that $\varepsilon_{a\tau_z}$ and 
$\varepsilon_{\mathrm A}$ decrease monotonically.  One can rationalize this finding by noting
that strong relaxation broadens the double-dot transition  and thereby suppresses coherent exchange processes between the cavity and the double dot. This effect reduces dynamical corrections beyond the leading dispersive approximation. The weak dependence of the phase error also confirms that the neglect of $\gamma$ in the derivation of Eq.~\eqref{weakcouplingA} from Eq.~\eqref{weakcouplingexactA} is justified for  
typical experimentally relevant parameters in the dispersive regime.

\new{
\subsubsection{Finite-temperature effects}\label{sec3b3}
}

\begin{figure}
    \centering
    \includegraphics[width=0.45\textwidth]{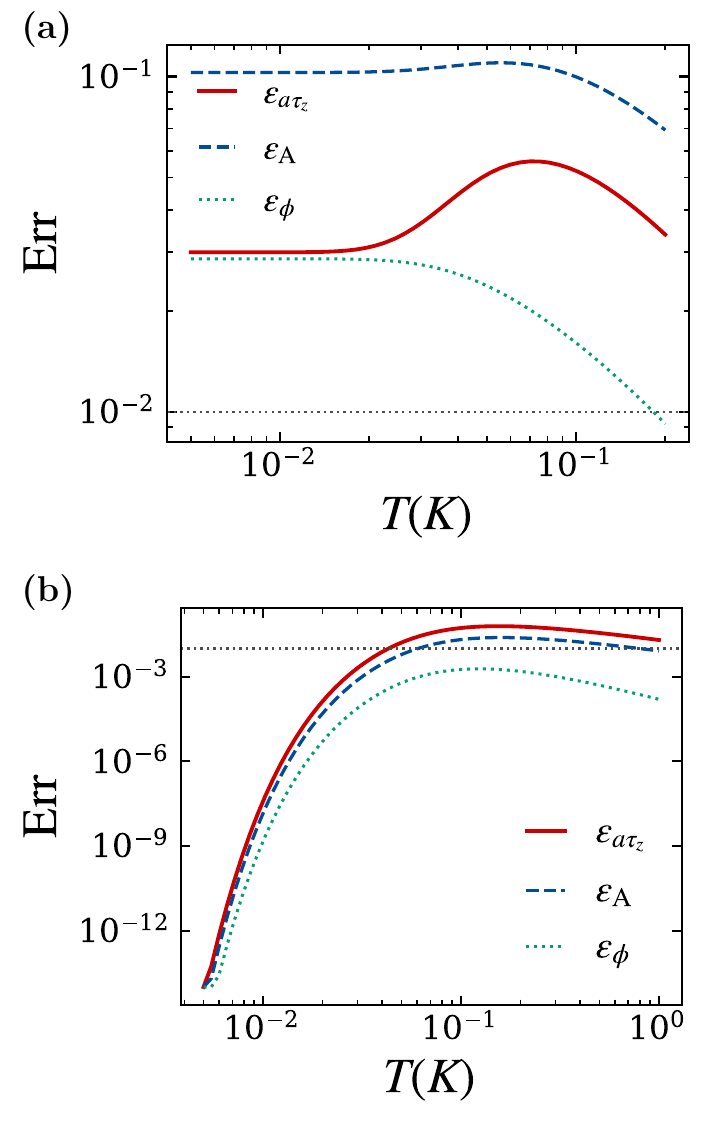}
    \caption{\new{Impact of thermal effects on different error measures for MBQ parity readout. Apart from temperature, the same model parameters as in Fig.~\ref{fig3} have been taken. The dotted horizontal line indicates an error threshold of $1\%$.
    (a) Strong-coupling regime, see Fig.~\ref{fig3}(a)--(d) for the corresponding low-temperature results. Raising the temperature leads to increased deviations from the semiclassical approximation, with a broad maximum in several error measures around $T\simeq 0.1$~K. 
    (b) Weak-coupling regime, see Fig.~\ref{fig3}(e)--(h).  We obtain  similar results as in the strong-coupling case, but the deviations from the semiclassical approximation now remain close to or even below the $1\%$ level at all temperatures studied.}}  \label{fig5}
\end{figure}

\new{
We finally turn to the impact of thermal effects on MBQ parity readout and the accuracy of the corresponding semiclassical expressions.  In order to study this question, we have performed additional simulations at elevated temperatures, see Fig.~\ref{fig5}.
Throughout this analysis, all parameters were kept fixed at the values used in Fig.~\ref{fig3} while temperature has been varied.
Figure~\ref{fig5}(a) shows the temperature dependence of the three error measures introduced in Eqs.~\eqref{var1}--\eqref{var3} in the strong-coupling regime.
At low temperatures, all these error measures approach constant values, indicating that here the dominant source for deviations from the semiclassical expression is not of thermal origin but rather due to the intrinsic breakdown of the semiclassical approximation caused by coherent photon-qubit correlations.
As temperature increases, the connected-correlator error $\varepsilon_{a\tau_z}$ develops a broad maximum around $T\simeq 0.1 \mathrm{K}$. 
This behavior arises from the thermal occupation of the excited double-dot state, which enhances photon-qubit correlations and thereby increases the deviations from the semiclassical approximation.
Interestingly, the amplitude error $\varepsilon_{\mathrm A}$ remains close to the $10\%$ level throughout the considered temperature range, indicating that the parity-dependent cavity response remains sensitive to correlation effects.
At even higher temperatures, all three error measures decrease again due to the thermal averaging of the double-dot state.
Indeed, as $\langle\tau_z\rangle_{\mathrm{ss}} \to 0$, the parity-dependent cavity response is progressively reduced, leading to a suppression of coherent photon-qubit correlations.}

\new{
The corresponding results in the weak-coupling limit are shown in Fig.~\ref{fig5}(b).
In contrast to the strong-coupling case, all error measures are strongly suppressed at low temperatures, taking values below the $1\%$ level at most temperatures.
As temperature increases, the errors measures grow and reach a broad maximum around $T\simeq 0.1$~K where they typically become of order $1\%$.
This increase is again driven by thermal excitations of the double dot.
Nevertheless, even near $T\simeq 0.1$~K, the semiclassical approximation remains rather accurate.  We thus conclude that semiclassical estimates can be used throughout the considered temperature range in the weak-coupling limit.}

\new{
In summary, the results in Fig.~\ref{fig5} show that thermal effects have the strongest impact if temperature becomes comparable to the double-dot level splitting.  (For the chosen parameters, this happens for $T\simeq 0.1$~K.)
In this case, the thermal occupation of the excited double-dot state enhances the photon-qubit fluctuations and thereby causes significant deviations from the semiclassical estimates.
By contrast, at much lower temperatures, the double dot remains predominantly in its ground state, limiting the impact of thermal excitations on MBQ parity readout.
For the parameters considered here, experimentally accessible temperatures $T\simeq 10$~mK stay well within this low-temperature regime.\\ 
}

\section{Conclusions}\label{sec4}

In this paper, we have studied charge reflectometry and capacitive readout of the parity degree of freedom associated with a Majorana pair in a Majorana box qubit.  The central quantity of interest in both readout schemes is the parity-dependent 
dynamical susceptibility $\chi_z(\omega)$.  We found a general expression for this quantity, see Eq.~\eqref{suscgen}, which covers the entire
crossover from the resonant strong-coupling regime to the off-resonant weak-coupling regime.  We have also critically re-examined 
previous results which were obtained from the Lindblad equation by making a semiclassical factorization assumption, see Eq.~\eqref{approx}.
In order to assess the accuracy of this approximation, we have compared the corresponding approximate results to exact reference results
by studying three different error measures as function of the system parameters, focusing on parameter regimes of current interest.
The reference results have been obtained either by a numerical solution for the steady state of the full Lindblad equation (in the strong-coupling case), or from exact analytical expressions (in the weak-coupling case). 
We find that in the weak-coupling regime, the semiclassical approximation provides a quantitatively accurate description for the cavity response, with errors staying below the $0.1\%$ level in all parameter regimes considered in this work.  
In the strong-coupling regime, however, results obtained under the semiclassical approximation typically deviate from the 
corresponding exact values by a few percent.   While such deviations are not expected to create major problems for parity readout experiments, they may play an important role if the contrast between both parity values is not pronounced.  In such cases, one should resort to a numerical solution of the full Lindblad equation.

\begin{acknowledgments} 
We thank K. Flensberg for discussions.
We acknowledge funding by the Deutsche Forschungsgemeinschaft (DFG, German Research Foundation) under Projektnummer 277101999 -- TRR 183 (project C01) and under Germany's Excellence Strategy -- Cluster of Excellence
Matter and Light for Quantum Computing (ML4Q) EXC 2004/2 – 390534769.\\
\end{acknowledgments}

\section*{Data availability}

The data underlying the figures in this work are available at Zenodo \cite{Zenodo}.

%

\end{document}